\documentclass[
reprint, prb,
aps,
superscriptaddress]
{revtex4-2}
\usepackage{graphicx} 
\usepackage{mathtools}
\usepackage{bm}
\usepackage{hyperref}
\hypersetup{
	linkcolor=black,
	colorlinks=true,
	urlcolor=blue,
	filecolor=black,
	citecolor=blue,
	pdftitle={Graphene Flower-Orbitals}
}
\date{\today}

\begin{document}
\title{Flower Orbitals in Graphene:\\
A Heavy-Fermion Lattice from a Gate-Defined Potential}

\author{Gerrit Niederhoff}
\affiliation{Institute of Physics, University of Zürich, 8057 Zürich, Switzerland}
\author{Benjamin Kilchmann}
\affiliation{Institute of Physics, University of Zürich, 8057 Zürich, Switzerland}
\author{Titus Neupert}
\affiliation{Institute of Physics, University of Zürich, 8057 Zürich, Switzerland}
\begin{abstract}

Strongly correlated and topological phases in graphene are frequently engineered through moiré superlattices of twisted multilayers.
Here we show that monolayer graphene alone, in the presence of a periodic electrostatic potential, is host to flat bands derived from  quasibound states of varying orbital angular momentum $m$.
These bands coexist with dispersive Dirac electrons and give rise to a family of strongly interacting models, selected via $m$: for large $m$, the effective theory is a Falicov-Kimball model, while for small $m$ a topological heavy-fermion model supports mean-field bands with nonzero Chern numbers. With this tunability, monolayer graphene joins the van der Waals material platforms that can be explored as quantum simulators.

\end{abstract}
\maketitle

\textit{Introduction ---}
Bound states in potential wells are elementary to quantum mechanics, and those in periodic potentials the core of Bloch's band theory of solids. With a surge of artificially engineered quantum systems, it is, however, ironic that the electrons in one of their best building blocks -- graphene -- do not admit bound states due to their relativistic nature. Rather, due to Klein tunneling, Dirac electrons in a potential well (or quantum dot) form \emph{quasibound} states  which can be understood as whispering gallery modes and coexist with a continuum of delocalized electrons. 
These states have been studied using numerical~\cite{fockdarwinGraphene}, analytical~\cite{analytialQuantumDot,quasiboundGraphene} and semiclassical ~\cite{fockdarwinGraphene,quasiRydbergGraphene,grapheneQuantumDots} methods. Scanning tunneling experiments have confirmed their existence in individual quantum dots~\cite{experimentalDOS,experimentaldIdV}. Their arrangement in lattices was studied theoretically, and peaks in the density of states of such a lattice were found corresponding to the quasibound states of the individual quantum dots.~\cite{quantumDotLattice}

Meanwhile, superlattice physics in graphene came from a different angle: the creation of moiré heterostructures~\cite{vdW,odedFlatbands,odedTopology}, most elementary in twisted bilayer graphene (TBG)~\cite{BMgraphene,tbg2}, allows for the precise creation of tunable superlattices. Different from quantum dots, these generate modulated hopping rather than a scalar potential. Moiré graphene systems have been shown to host a phenomenology that uniquely combines topological band structures with strong correlation physics. For instance, the flatband manifold of TBG -- with spin, flavor and valley degeneracy -- can be decomposed into sectors with opposite-in-sign Chern numbers. Interactions consequently select anomalous quantum Hall states spontaneously at odd-integer fillings~\cite{grapheneCI3/4filling,grapheneCorrelatedInsulatorsandCIs,grapheneCIexperiment,grapheneCIexperiment2}, while more intricate orders are seen at even integer fillings~\cite{Kekule1,Kekule2}. Doping these mean-field states gives rise to superconductivity and fractional Chern insulators~\cite{CaoSuperconductivity,KhalafTBGSC,XieFCI,TitusFCI,RegnaultFCI, ShengFCI,TBGFCItheory}. 
Complementarily, TBG admits a heavy-fermion description in terms of a periodic Anderson model where  `$f$-electron' Wannier orbitals coexist with Dirac electrons and hybridize in a small portion of the Brillouin zone, which gives rise to band topology.~\cite{HeavyFermions1}

In this work, we revisit quasibound states in graphene quantum dots, 
extend them to quantum dot lattices, 
and study them from the perspectives developed in TBG: flat bands with quantum geometry and topology and the separation into an atomic limit of local moments and itinerant electrons. 
We find that individual quantum dots can host a tower of sharp quasibound states that are organized by angular momentum $m$ and principal (radial) quantum numbers $n$, coming in degenerate pairs of $\pm m$. These states -- the \emph{flower orbitals} -- can be found semiclassically and have support in a ring defined by the potential parameters. States with higher radial quantum numbers have shorter lifetimes.

We proceed by considering a triangular lattice of such quantum dots.

Numerical diagonalization of the resulting Hamiltonian reveals sequences of flat bands coexisting with Dirac electrons. The ratio of bandwidth/hybridization to separation between flat bands gets as low as $10^{-2}$ and is highly tunable via their quantum numbers. While these flat bands share their eightfold degeneracy (angular momentum, spin, valley) with the magic angle flat bands of TBG, they differ in that they appear at finite energies and in hierarchical sequences. 

Interactions can famously lift flat band degeneracies to give rise to various correlated states.~\cite{flatbandKondoReviewColeman,TitusFCI,RegnaultFCI,ShengFCI,flatbandSC,CaoCorrelated,CaoSuperconductivity} Together with the tunability of van der Waals materials, this allows one to consider them as \emph{quantum simulators} supporting paradigmatic strongly correlated electron models~\cite{kennesSimulatorReview}. In this spirit, we use Hartree-Fock calculations to demonstrate that flower orbital flat bands can host a family of models, where $m$ takes the role of the main design parameter (instead of, e.g., the twist angle). In the limit of large $m$, the system generically realizes Falicov--Kimball physics~\cite{FalicovKimball}, while for small $m$ the flat and dispersive bands hybridize into interaction-driven Chern insulators, akin to the topological heavy-fermion phases of TBG.

\begin{figure*}
    \centering
    \includegraphics[]{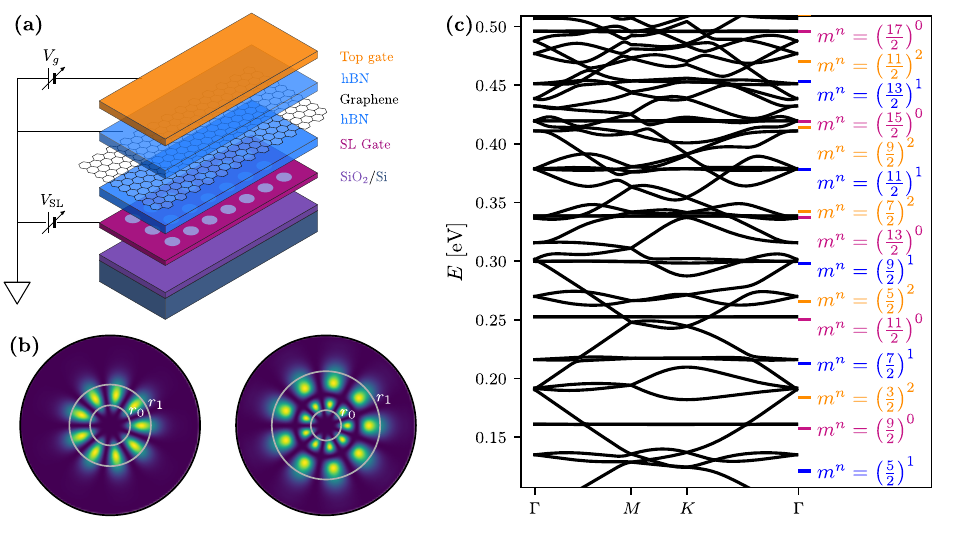}
    \caption{(a) Schematic experimental set-up that allows for control of the superlattice potential $V_\text{SL}$ as well as tuning the chemical potential using $V_g$. The schematic is an adaptation from Ref.~\onlinecite{CanoFlatbands}
    (b)
    Absolute value squared of the first spinor-component of the $m=\pm\frac92$ quasibound wave-function, ground state (left) and first excited state (right) in the linear potential well.
    (c)
    Bandstructure of the potential lattice model together with the semiclassical predictions of the energies of $n$th excited quasibound states with angular momentum $\pm m$. The labels and lines are color coded for better readability, with ground states $n=0$ in violet, $n=1$ in blue and $n=2$ in orange.
    }
    \label{fig1}
\end{figure*}

\textit{Quasibound states in quantum dots ---}
 Throughout this work, we focus on electrons in graphene subject to a spatially modulated electrostatic potential.
Experimentally, this can be achieved by stacking hexagonal boron nitride (hBN) encapsulated monolayer graphene on top of a patterned dielectric~\cite{patternedDielectricExperiment}, as shown in Fig.~\ref{fig1}(a), with top and back gates added. The top gate voltage $V_g$ controls the uniform doping, while the back gate $V_{\text{SL}}$ imprints and controls the superlattice (SL) potential.
 
We first focus on a dielectric with an isolated quantum dot potential.
We assume that the characteristic potential length-scale is much larger than the inter-atomic distances in graphene, so that the electrons can be treated in the standard continuum approximation.  The full non-interacting model has spin-rotation symmetry and no intervalley coupling, hence we do not explicitly consider these degrees of freedom.

The problem of 2D Dirac electrons in a central potential is not analytically solvable in general. It generically admits no true bound state solutions, only quasibound states, which can be investigated numerically or semiclassically~\cite{fockdarwinGraphene,grapheneQuantumDots,quasiboundGraphene,quasiRydbergGraphene}, or in some cases analytically~\cite{analytialQuantumDot}.
In this work, we derive a semiclassical solution, assuming a rotationally symmetric potential. 
With this, the entire Hamiltonian is invariant under rotations and can be solved with (half-integer) total angular momentum $m$ as a good quantum number, with the remaining degree of freedom being described by a radial wave-number $k_r$. The details of the semiclassical calculation can be found in the appendix. There, we diagonalize the Hamiltonian
\begin{align}
    H &= V(r)+
    \hbar v_\text{F}
    \begin{pmatrix}
        V(r)&\hat{k}_x-i\hat{k}_y\\
        \hat{k}_x+i\hat{k}_y&V(r)
    \end{pmatrix},
\end{align}
with $v_\text{F}$ the Fermi velocity in graphene, such that $\hbar v_\text{F}\approx0.6582$ eV$\cdot$nm. We then
 fix an energy $E$, locally solve for $k_r$, and  impose a constructive interference condition (Bohr-Sommerfeld quantization) on the integrated momentum as the particle travels between two boundaries of a classically allowed region. These allowed regions are defined by demanding that $k_r$ be real. 
Imposing Bohr-Sommerfeld quantization yields the following equation:
\begin{align}
    2\int_{r_0}^{r_1}\text{d}rk_r + \theta_\text{R} = 2\pi n,\label{quantizationCondition}
\end{align}
where $\theta_\text{R}$ is a Maslov correction, a phase picked up on reflection. In the appendix, the phase upon a single reflection is found to be $\frac\pi2$, making $\theta_\text{R}=\pi\, \mathrm{mod}\, 2\pi$, which agrees with previous work~\cite{grapheneQuantumDots}. 
For a simple linear potential $V(r)=\lambda r$, where we used $\lambda\approx 74\frac{\text{meV}}{\text{nm}}$, the first spinor-component density is plotted in Fig.~\ref{fig1}(b) for $m=\pm\frac92$. To clearly resolve the angular structure in the absolute value of the wavefunction, we plot an equal weight superposition $|\Psi\rangle = \frac{1}{\sqrt{2}}\left(|m=\frac92\rangle+|m=-\frac92\rangle\right)$. The angular shape of these wave-functions is what leads us to calling them flower orbitals, a name which has previously been used for similarly shaped wavefunctions in graphene~\cite{odedFlatbands}.

Figure~\ref{fig1}(b) shows the confinement to the classically allowed region, the angular structure of 9 peaks, and the radial structure where a node is added in the first excited state. 
The tunneling amplitude through the classically forbidden region is a proxy for the lifetime of the quasi-bound states. This calculation has been the subject of previous works.~\cite{fockdarwinGraphene}\\

\textit{Flat bands in a periodic potential ---}
We now turn to a periodic lattice of quantum dots, whose spectrum can be described through its band structure. The localized quasi-bound states of the individual dots then appear as (nearly) flat bands.
To facilitate solving for the spectrum in momentum space, following the work of Bistritzer and MacDonald~(BM) for TBG~\cite{BMgraphene}, we focus on potentials with low-order harmonics of some Bravais lattice. One retains reciprocal lattice sites up to some UV-cutoff, and then diagonalizes the Hamiltonian numerically. 

We seek a potential with a high degree of rotational symmetry around extrema of the potential such that a connection with our semiclassical calculation for an individual quantum dot can be made.

A $C_6$-symmetric potential meets this requirement best, as there the leading anisotropy around potential extrema is of sixth order. We thus consider a triangular lattice potential built from the lowest order harmonics 
\begin{align}
    V(\bm{r})
    &=V_\text{SL}
    \sum_{j=1}^3
    e^{i\bm{q}_j\cdot\bm{r}}+\text{c}.\text{c}.\label{potentialform},
\end{align}
where we choose $V_\text{SL}=-0.11$ eV, and $\bm{q}_1=q_0 (0,1)^{\mathrm{T}}$, $\bm{q}_2=q_0(\sqrt{3},1)^{\mathrm{T}}/2$, and $\bm{q}_3=q_0(-\sqrt{3},1)^{\mathrm{T}}/2$. Here, $q_0$ determines the length of the real space lattice vectors according to $|\bm{R}_i|=\frac{2}{\sqrt{3}}\frac{2\pi}{q_0}$, we choose $q_0$ such that $|\bm{R}_i|\approx33.9$ nm.~\footnote{For reference: This length scale corresponds to a twist angle in TBG of $\theta=0.72^\circ$} 

The band structure for this potential is shown in Fig.~\ref{fig1}(c), juxtaposed to the spectrum of an individual quantum dot within semiclassical analysis. For the latter, we average the potential around a minimum over all angles to impose rotation symmetry.  

The distinctive feature of the band structure in Fig.~\ref{fig1}(c) is a sequence of flat bands, whose energies are in good correspondence to those of quasi-bound states of the individual quantum dot, obtained semiclassically. The shorter lifetime of the states with higher radial quantum number manifests itself in a bigger bandwidth. Most of the flatbands show virtually zero hybridization with the Dirac cone-like dispersive bands that cross them. We attribute this to the fact that the very large difference in angular momentum between the flower orbital and the Dirac cone suppresses any transitions between the states. Generally, hybridization is not strictly forbidden since the crossings do not occur at high symmetry points. Along mirror lines the degenerate flat-bands will have mirror eigenvalues $\pm1$, which means that at exactly one of them is allowed to hybridize with the dispersive band, which will have either $+1$ or $-1$ mirror eigenvalue. Scaling $\bm{q}_i$ and $V_\text{SL}$ by the same factor $s$ leaves the shape of the spectrum invariant, but rescales it by $s$.

In Ref.~\onlinecite{CanoFlatbands}, Cano et al.\ considered the same triangular potential imposed on Bernal bilayer graphene, similarly finding a sequence of flat bands.

However, 
these flat bands appear spectrally isolated instead of coexisting with Dirac electron bands.
An experimental study on the systems described by Cano et al., evidence for correlated insulating states, consistent with their predictions was found.~\cite{CanoExperiment}

There have also been experiments with the exact setup we propose for our model, though these were not focused on finding evidence of flat bands or correlated insulators.~\cite{patternedDielectricExperiment}

\textit{Interactions and topology ---}
Electron-electron interactions will significantly renormalize the electronic structure in the presence of flat bands and may induce strongly correlated states. Here, we concern ourselves with the former effect to study (i) the spontaneous lifting of the flower orbital degeneracy, (ii) interaction-induced hybridization between Dirac band and flat bands, and (iii) thereby induced band topology.
We use self-consistent Hartree-Fock (HF) mean-field calculations projected onto a small number of relevant bands at zero temperature, closely following Ref.~\onlinecite{meanfieldNecronomicon}. 

We consider a dual gate-screened Coulomb interaction $V_q = \frac{e^2}{2\epsilon_0\epsilon_rq}\tanh{qd_{\text{sc}}}$, with a relative permittivity $\epsilon_r=9$ and a screening length $d_\text{sc}=30$ nm (unless otherwise stated), both of which are within a realistic range for graphene devices~\cite{meanfieldNecronomicon}. Furthermore, HF calculations were done using the ``average" scheme for a reference state~\cite{meanfieldNecronomicon}, and at half filling of the flat-band manifold. Chern numbers were then calculated using the Fukui-Hatsugai-Suzuki (FHS) method~\cite{FHS-chernnumber}, see the appendix for details.

\begin{figure*}
    \centering
    \includegraphics[]{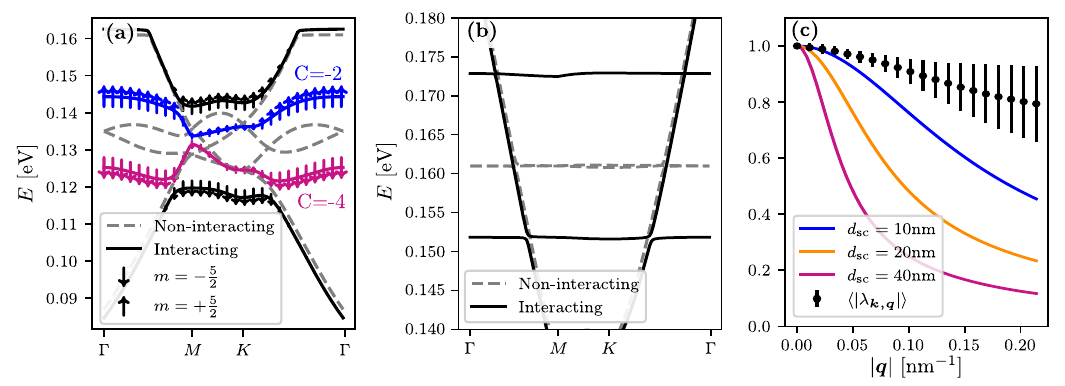}
    \caption{(a)
    Mean-field and non-interacting band structure for an $m=\pm\frac52$ flower band, with highlighted bands having Chern numbers $C=-4$ and $C=-2$ as indicated. The length of the arrows pointing up (down) is proportional to the overlap with the $m=+\frac52$ ($m=-\frac52$) flower orbital.
    (b)
    Mean-field and non-interacting band structure for an $m=\pm\frac92$ flower band. 
    (c)
    Variation of form factor $|\lambda^{\tau\tau}_{\bm{k},\bm{q}}|$ in $\bm k$, together with the screened Coulomb interaction $V_{\bm{q}}$ for different values of the screening length $d_{\text{sc}}$ (all normalized by the respective maximum value).
    }
    \label{fig2}
\end{figure*}

The nature of the interaction-induced renormalization of a flat band is primarily determined by the angular momentum $m$ of its flower orbitals. For large $m$ the bands remain nearly unhybridized, while for small $m$ substantial hybridization leads to potentially topological (partially) flat bands. Studying flat bands of varying $m$ allows one to tune between these limits, which we refer to as the \emph{Falicov-Kimball limit} (large $m$) and the \emph{topological heavy-fermion limit} (small $m$), respectively. The hybridization should thereby be compared to energy scales of temperature and disorder, which determine whether it results in observable effects.
In all cases, the $\pm m$ degeneracy is spontaneously lifted, giving rise to orbital angular momentum ferromagnets. 

We now discuss these two limiting cases.
In the topological heavy-fermion limit, interactions induce hybridization of the flower orbitals with the Dirac cones and open a topological gap. 
The mechanism by which the hybridized bands acquire a Chern number $C$ is a higher-angular-momentum variant of the band inversion of topological insulators. 
Instead of an inversion symmetry based topological indicator, the product of rotation eigenvalues at different high-symmetry momenta constrains $C$, in case at hand to $C\,\mathrm{mod}\,6$ due to the sixfold rotation symmetry~\cite{FangTopology}.
The flower orbitals form trivial atomic-limit bands with $C_6$ eigenvalues $e^{-2\pi i(\pm m+1/2)/6}$ at $\Gamma$, and $e^{-2\pi i(\pm m+1/2)/3}$ and $e^{-\pi i(\pm m+1/2)}$ for the $C_3$ and $C_2$ eigenvalues at $K$ and $M$, respectively. 
The folded Dirac bands realize different band representations in general, so that hybridization with the flower-orbital bands can exchange eigenvalues at one high-symmetry point, and thereby enforce nonzero Chern numbers [see appendix,  Eq.~\eqref{bandinversion}].

An example of a topology driven by interactions in this limit is shown in Fig.~\ref{fig2}(a) for the $m=\pm\frac52$ flower orbital. Here, the two central bands exchange their rotational eigenvalues at $K$ and $M$, visualized by the arrows showing the angular momentum character of the band. This fixes $C=2$ and 4 ($\mathrm{mod}\, 6$), consistent with $C=-4$ and $-2$ obtained from integrating the Berry curvature  via the FHS formula (see appendix). Note that, as can be seen from the non-interacting band-structure in Fig.~\ref{fig2}(a), this flower orbital band did not hybridize with the crossing Dirac band in the bare band structure, even though it is more dispersive than its cousins of higher angular momentum: The gap opening is purely an interaction effect.

In contrast, the Falicov-Kimball limit is characterized by the absence of hybridization between flat bands and Dirac electrons on the Hartree-Fock level, a relativistic realization of the Falicov-Kimball model~\cite{FalicovKimball}.  
An example is the $m=\pm\frac92$ flower orbital, for which the HF bands are shown in Fig.~\ref{fig2}(b).
The effective non-interacting part of the Hamiltonian is
\begin{align}
    H^0_\text{eff} =-\sum_{\bm{k}\sigma}
    \hbar v_\text{F}|\bm{k}|
    d^\dagger_{\sigma\bm{k}}d_{\sigma\bm{k}}
    -E_0
    \sum_{\bm{k}\tau}f^\dagger_{\tau\bm{k}}f_{\tau\bm{k}},\label{effectivemodel}
\end{align}
with the $d$ operators corresponding to the dispersive Dirac electrons and the $f$ operators to the localized states in the flower orbitals. The band index $\sigma$ includes spin and valley degeneracy for the $d$-operators, and $\tau$ spin, valley, and orbital angular momentum ($\pm m$ for a fixed $m=|m|$) degeneracy for the $f$-operators.
The defining property of a Falicov-Kimball model is that the density operators of the $f$ electrons, $\hat{n}_{\tau\bm{R}}$, are conserved quantities at every site (i.e., superlattice unit cell) $\bm{R}$, where $\hat{n}_{\tau\bm{q}}\coloneqq\sum_{\bm{k}}f^\dagger_{\tau\bm{k+q}}
f_{\tau\bm{k}}=\sum_{\bm{R}}e^{-i\bm{q}\bm{R}} 
\hat{n}_{\tau\bm{R}}$.
In the effective interacting Hamiltonian, obtained from projecting the Coulomb interaction onto the $d$ and $f$ electron bands, the projected density operator generically takes the form $\sum_{\bm k}\lambda^{\tau\tau'}_{\bm k,\bm q} f^\dagger_{\tau\bm k+\bm q} f_{\tau'\bm k}$, involving the form factors  defined as $
    \lambda^{\tau\tau'}_{\bm{k},\bm{q}}
    \coloneq
    \langle u^\tau_{\bm{k+q}}|u^{\tau'}_{\bm{k}}\rangle$ with 
$|u^\tau_{\bm{k}}\rangle$ being the cell-periodic parts of the Bloch states (similarly we define $\lambda^{\sigma\sigma'}_{\bm{k},\bm{q}}$ form factors for the $d$ bands and $\lambda^{\sigma\tau}_{\bm{k},\bm{q}}$ form factors connecting $f$ and $d$ bands)~\cite{meanfieldNecronomicon}. This projected density reduces to an $\hat{n}_{\tau\bm{R}}$-conserving form iff 
\begin{equation}
    \lambda^{\tau\tau'}_{\bm k,\bm q} = \delta_{\tau\tau'}F_{\tau\bm q}
    \label{eq:condition conserved f density}
\end{equation}
is independent of $\bm{k}$, where $F_{\tau\bm q}$ is the Fourier transformed flower orbital density that can be absorbed into the bare interaction $V_{\bm q}$. At small $\bm q$, Eq.~\eqref{eq:condition conserved f density} implies a $\bm k$-independent quantum metric of the flower-orbital bands.

The Falicov-Kimball Hamiltonian is obtained if (i) Eq.~\eqref{eq:condition conserved f density} is satisfied, (ii) the form factors $\lambda^{\sigma\tau}_{\bm{k},\bm{q}}$ between $f$ and $d$ vanish, and (iii) the $f$ band energy is $\bm k$-independent. We find all of these properties hold to a good approximation for flat bands with high $m$. We exemplify this for $m=9/2$, for which the Hartree-Fock band structure is shown in Fig.~\ref{fig2}(b). Specifically, the variation of $\lambda^{\tau\tau'}_{\bm k,\bm q}$ in $\bm k$, shown in Fig.~\ref{fig2}(c), is negligible only for $|\bm q|<1/R$, with $R$ the flower orbital radius.
In summary, the interaction part of the effective Hamiltonian reads in the Falicov-Kimball limit
\begin{align}
\begin{split}
    H^{ff}_\text{int} &= \frac12\sum_{\tau, \tau',\bm{q}}V_{\bm{q}} F_{\tau\bm q} F_{\tau'\bm -q}
    \hat{n}_{\tau\bm{q}}\hat{n}_{\tau'\bm{-q}},\\
    H^{fd}_\text{int}&=
    \frac12\sum_{\tau,\bm{q}}
    V_{\bm{q}} F_{\tau\bm q} \hat{n}_{\tau\bm{q}}
    \sum_{\sigma\bm{k}}
    \lambda^{\sigma\sigma}_{\bm{k+q},\bm{-q}}
    d^\dagger_{\bm{k}\sigma}
    d_{\bm{k+q}\sigma}.
\end{split}\label{effectiveInteraction}
\end{align}
 At commensurate filling, the dominant $ff$-interaction forces the system into a ground state with one of the $\pm m$  flower orbitals occupied per superlattice unit cell, seen in the splitting of the flat bands in Fig.~\ref{fig2}(b).

\textit{Conclusions and outlook --}
We showed that graphene subject to a periodic potential hosts flat bands that are distinct in origin from those in moiré superlattices and admit an effective description tunable between a Falicov-Kimball limit and a topological heavy-fermion limit. The latter features bands with nonzero Chern numbers, akin to the hybridization-gap mechanism of topological Kondo insulators. 

Notably, the coexisting localized and itinerant degrees of freedom emerge microscopically from the same electrons -- populating graphene $p_z$ orbitals -- in contrast to functionalized graphene systems~\cite{functionalizedGraphene}. The emergent picture is thus somewhat  reminiscent of ancilla-layer descriptions of the cuprates~\cite{cuprateAncillaQbits}, with the difference that here the localized states are physical, charged eigenstates of the Dirac problem rather than auxiliary spins.

Aside from motivating the experimental investigation of this system, our study opens natural avenues for future theoretical work. For instance, while the conventional Falicov–Kimball model has a famously rich phase diagram, its relativistic counterpart remains largely unexplored.

\bibliography{main}
\appendix
\onecolumngrid
\section{Semiclassical Calculation}\label{semiclassicaldetails}
The Hamiltonian of interest is Eq.~\eqref{polarHamiltonian}.
 From now on, we set $\hbar v_\text{F}=1$. This Hamiltonian commutes with the following (projective) representation action of SO(2)
\begin{align}
    R_\theta \coloneqq R_\theta^\text{spinor}sR_\theta^\text{orbital},\qquad
    R_\theta^\text{spinor} \coloneqq \exp\left(-i\frac\theta2\sigma_z\right),
    \qquad
    R_\theta^\text{orbital} = \exp\left(-i\theta\partial_\varphi\label{rotationoperator}\right)
\end{align}
where $\varphi$ is the polar angle.
This treats the models sublattice degree of freedom like a physical spin-$\frac12$. One could include another overall phase in the rotation to change this, but that changes nothing of the physics, and the calculations are simpler in this form. Since this means that the system is invariant under total angular momentum (meaning orbital angular momentum and pseudo-spin), it can be solved in the half-integer angular momentum eigenspaces. We use polar, real-space coordinates, and the basis states
\begin{align}
    \left|m;1\right\rangle &= e^{im\varphi}\begin{pmatrix}
        e^{-i\varphi/2}\\
        0
    \end{pmatrix},
    \qquad
    \left|m;2\right\rangle = e^{im\varphi}\begin{pmatrix}
        0\\
        e^{+i\varphi/2}
    \end{pmatrix},
\end{align}
where $m$ is the eigenvalue of the angular momentum operator $J_z\coloneqq \frac12\sigma_z-i\partial_\varphi$. The rotation operator around the $z$-axis (see Eq.~\eqref{rotationoperator} acts on the basis vectors $|m;1\rangle$ and $|m;2\rangle$ as 
\begin{align}
    R_\theta|m;1\rangle &= e^{im(\varphi-\theta)}\begin{pmatrix}
        e^{-i\varphi/2}\\0
    \end{pmatrix},\qquad
    R_\theta|m;2\rangle = e^{im(\varphi-\theta)}\begin{pmatrix}
        0\\e^{i\varphi/2}
    \end{pmatrix}.
\end{align}
In this basis, the Hamiltonian takes the form
\begin{align}
  \widetilde{H} = \begin{pmatrix}
    V(r) & k_r-i \frac{m}{r}\\
    k_r +i \frac{m}{r} & V(r)
  \end{pmatrix}\label{polarHamiltonian},
\end{align}
with the radial momentum operator $k_r\coloneqq -i\partial_r-\frac{i}{2r}$.

Note that the Hamiltonian is also invariant under a pseudo-time-reversal symmetry $\mathcal{T}\coloneqq i\sigma_y\mathcal{K}$ (again, treating the Dirac pseudo-spin as a physical spin), which guarantees that the solutions with $\pm m$ will be degenerate in each valley. This does not correspond to a symmetry of the entire system, and is only valid when the $K$ and $K'$ point can be taken as truly independent~\cite{pseudoTRS} and without trigonal warping.

Diagonalizing the Hamiltonian in Eq.~\eqref{polarHamiltonian} yields the eigenvalues and eigenvectors
\begin{align}
  E^\pm_m = V(r) \pm \sqrt{\frac{m^2}{r^2}+k_r^2},\qquad u^\pm_m(r,k_r) &= \begin{pmatrix}
    1\\\pm e^{i\alpha_{r,m}}
  \end{pmatrix},\label{semiclassicalstates}
\end{align}
where $\alpha_{r,m}=\arctan{\frac{m}{rk_r}}$. 
Solving for $k_r$ in the eigenvalues Eq.~\eqref{semiclassicalstates} yields the expression:
\begin{align}
  k_r = \sqrt{(E-V(r))^2-\frac{m^2}{r^2}},
\end{align}
which can be used to write an ansatz wavefunction in the region $r_0<r<r_1$, defined as the region where $k_r$ is real. In the classically forbidden region, an imaginary radial momentum should be used, making the overall Ansatz~\cite{grapheneOptics}
\begin{align}
    \psi_1(r) &= 
  e^{im\varphi}\left[
    \exp\left({i\int_{r_0}^{r}k_r\text{d}r}\right)\begin{pmatrix}
      1\\e^{i\alpha_r}
    \end{pmatrix}+r_f
    \exp\left({-i\int_{r_1}^{r}k_r\text{d}r}\right)\begin{pmatrix}
      1\\e^{i(\pi-\alpha_r)}
    \end{pmatrix}
  \right]
  \\
  \psi_2(r) 
  &=
  \left[
    A\exp\left(-\int_{r_1}^{r} \rho_r \text{d}r dx\right)
    \begin{pmatrix}
      1\\i\sqrt{\frac{m+r\rho_r}{m-r\rho_r}}
    \end{pmatrix}
  \right],\qquad
  \rho_r = \sqrt{\frac{m^2}{r^2}-(E-r)^2},
\end{align}
Note that the phase $\pi-\alpha_r$ in the reflected wave is because that wave has negative radial momentum (but $k_r$ is taken to be positive everywhere, so that it is single-valued). Writing the phase like this ensures that the phase of the spinor varies continuously as $k_r$ is tuned through 0.\\
The reflection phase can be deduced by determining the value of $r_f$, which we do by demanding continuity (and then also continuity of the derivative). Defining $I\coloneq\int_{r_0}^{r_1}k_r\text{d}r$, the two equations we get from demanding a continuous wave-function at $r_1$ are
\begin{align}
    \exp{iI}+r_f&=A\\
    i\exp{iI}+ir_f&=iA
\end{align}
The fact that the two spinor-components yield the same equation is why we need to demand a continuous derivative as well, which otherwise isn't necessary for Dirac electrons~\cite{grapheneOptics}. The derivatives of the exponentiated integrals vanish at the reflection points, since these are by definition points where $k_r$ and $\rho_r$ vanish. The derivatives of the second spinor components of both $\Psi_1(r)$ and $\Psi_2(r)$ actually diverge at these reflection points, but they diverge in the same way, so the limit of their ratio is well-defined, which allows us to still match the coefficients. In fact, assuming $V(r)$ around the reflection point can be linearised as $V(r)\approx V(r_1)+V'|_{r_1}\cdot(r-r_1)$, we get
\begin{align}
    \lim_{\delta\to0}\left(
    \frac{\left.rk_r\right|_{r_1-\delta}}
    {\left.r\rho_r\right|_{r_1+\delta}}
    \right)=
    \lim_{\delta\to0}\left(
    \frac{\left.\partial_r(rk_r)\right|_{r_1-\delta}}
    {\left.\partial_r(r\rho_r)\right|_{r_1+\delta}}
    \right)
    =
    -1,\label{symmetricLimit}
\end{align}
which helps after writing both relevant terms in terms of $y\coloneqq rk_r$ and $y'=r\rho_r$. We have 
\begin{align}
    \alpha_r &= \arctan\left(\frac{m}{y}\right),
    \qquad
    \partial_r\alpha_r =
    \frac{-m}{m^2+y^2}\partial_ry\\
    \partial_r\sqrt{
    \frac{m+y'}{m-y'}
    }
    &=
    \sqrt{\frac{m-y'}{m+y'}}\frac{m}{m-y'^2}\partial_ry'
\end{align}
The condition for a 'continuous' derivative from our Ansatz is
\begin{align}
    i^2e^{iI}\partial_r\alpha_r|_{r=r_0}
    -r_fi^2\partial_r\alpha_r|_{r=r_0}
    &=
    iA\partial_r\left.\sqrt{
    \frac{m+y'}{m-y'}
    }\right|_{r=r_1}\\
    \Longleftrightarrow\quad
    e^{iI}\frac{\partial_ry|_{r=r_1}}{m}
    -r_f\frac{\partial_ry|_{r=r_1}}{m}
    &=
    iA\frac{\partial_ry'|_{r=r_1}}{m}.
\end{align}
At $r=r_1$, we can identify $y$ and $y'$ and their derivatives according to Eq.~\eqref{symmetricLimit} which gives us a second equation relating $r_f$ and $A$:
\begin{align}
    e^{iI}-r_f=-iA.
\end{align}
Combining this with the continuity condition yields
\begin{align}
    e^{iI}-r_f &= -ie^{iI}-ir_f \\
    \Longleftrightarrow
    r_f &= \frac{1+i}{1-i}e^{iI}=e^{i(I+\frac\pi2)}
\end{align}
The phase $I$ just confirms the obvious intuition that the wave retains the phase it picked up while traveling through the allowed region, the $\frac\pi2$ is a phase jump due to reflection. Given that a wave travelling back and forth is reflected twice before repeating its path, this phase is picked up twice (note that we showed the analysis only at $r_1$ here, but the calculation at $r_0$ is completely the same and yields the same phase). This gives the overall additional phase of $\theta_\text{R}=\pi$ that goes into the quantization condition Eq.~\eqref{quantizationCondition}.

\section{Calculating Chern Numbers} 
\begin{figure}
    \centering
    \includegraphics[]{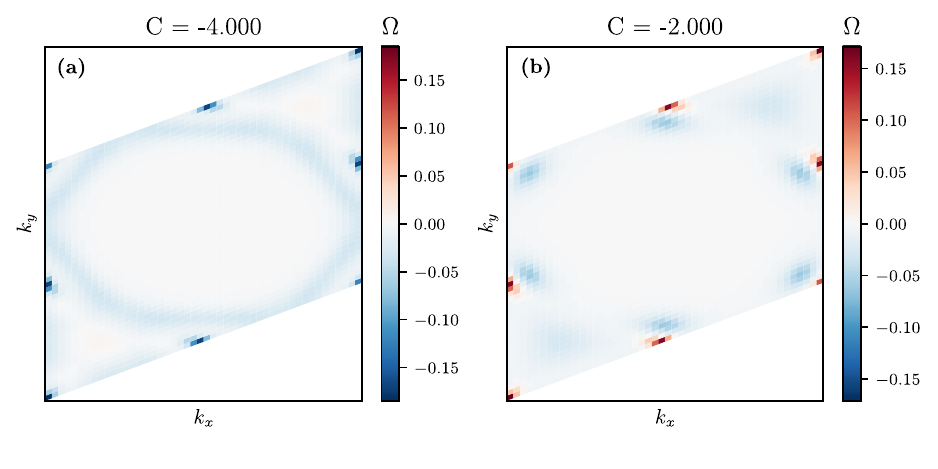}
    \caption{Berry curvatures for the two Chern-bands in Fig.~\ref{fig2}(a), the lower band in (a) and the upper band in (b). The color shows the dimensionless Berry-flux going through each plaquette, which is defined modulo $2\pi$.}
    \label{chernfigure}
\end{figure}
Here, we give some more details regarding our arguments for the non-zero Chern numbers indicated in Fig.~\ref{fig2}(a). The formula by Fang et al.~\cite{FangTopology} we refer to for systems with $C_6$ symmetry is
\begin{align}
    e^{C2\pi i/6} &= \prod_{i\in\text{occ.}} (-1)^F \eta_i(\Gamma)\theta_i(K)\zeta_i(M)\label{fangformula}
\end{align}
where $i$ enumerates the occupied bands and $\eta_i(\Gamma)$, $\theta_i(K)$, and $\zeta_i(M)$ represent the $C_6$, $C_3$ and $C_2$ eigenvalues at the respective invariant $\bm{k}$-points. $F$ is twice the total spin of the particles, so 0 in our case, considering spinless fermions. Since we take $F=0$ for this, we multiply the spinor part of the rotation operator by an overall phase, $\tilde R^{\rm spinor}_\theta := e^{-i\theta/2}R^{\rm spinor}_\theta$, 
\begin{align}
    \widetilde{R}^\text{spinor}_\theta \coloneqq \exp\left(-i\frac\theta2\sigma_z-i\frac\theta2\right),
\end{align}
such that $\tilde R_{2\pi}=1$. The eigenvalues for the $\pm\frac52$-flower orbital at the high symmetry points are
\begin{align}
\begin{split}
    \eta_{-\frac52}(\Gamma) &= e^{\frac{2\pi}{3}i},
    \qquad
    \theta_{-\frac52}(K) = e^{\frac{4\pi}{3}i}
    ,\qquad
    \zeta_{-\frac52}(M) = e^{0i},\\
    \eta_{\frac52}(\Gamma) &= e^{\pi i},
    \qquad
    \theta_{\frac52}(K) = e^{0 i}
    ,\qquad
    \zeta_{\frac52}(M) = e^{\pi i},
\end{split}
\end{align}
and the Dirac cone crossing at $K$ and $M$ turns out to have the same eigenvalues there. Note that a band with rotation eigenvalues belonging to the same angular momentum at all high-symmetry points will be trivial. The mean-field Chern bands shown in Fig.~\ref{fig2}(a) then turn out to have eigenvalues
\begin{align}
\begin{split}
    \eta_{C=-4}(\Gamma) &= e^{\frac{2\pi}{3}i},
    \qquad
    \theta_{C=-4}(K) = e^{0i}
    ,\qquad
    \zeta_{C=-4}(M) = e^{0i},\\
    \eta_{C=-2}(\Gamma) &= e^{\pi i},
    \qquad
    \theta_{C=-2}(K) = e^{\frac{4\pi}{3}i}
    ,\qquad
    \zeta_{C=-2}(M) = e^{\pi i}.
\end{split}\label{bandinversion}
\end{align}
Note the inversion (swapping of eigenvalues), which forces the lower band to have $C\text{ mod }6=2\text{ mod }6=-4\text{ mod }6$ and the upper band to have $C\text{ mod }6=4\text{ mod }6=-2\text{ mod }6$. The Chern number was also computed as an integral over the Berry curvature using the standard FHS method for computing Chern numbers in a discretized BZ~\cite{FHS-chernnumber}. The Berry curvature for both bands is shown in Fig.~\ref{chernfigure}. Observe that the peaks in Berry curvature are located around the $M$-points, which is where the bands are closest to each other.
\end{document}